# Critical current and linewidth reduction in spin-torque nano-oscillators by delayed self-injection

Guru Khalsa,[a)] M. D. Stiles

Center for Nanoscale Science and Technology, National Institute of Standards and Technology, Gaithersburg, MD 20899-6202, USA

J. Grollier

Unité Mixte de Physique CNRS/Thales, 1 avenue A. Fresnel, Campus de l'Ecole Polytechnique, 91767 Palaiseau, France, and Université Paris-Sud, 91405 Orsay, France

**ABSTRACT:**

Based on theoretical models, the dynamics of spin-torque nano-oscillators can be substantially modified by re-injecting the emitted signal to the input of the oscillator after some delay. Numerical simulations for vortex magnetic tunnel junctions show that with reasonable parameters this approach can decrease critical currents as much as 25 % and linewidths by a factor of 4. Analytical calculations, which agree well with simulations, demonstrate that these results can be generalized to any kind of spin-torque oscillator.

Spin-torque nano-oscillators (STNOs) based on magnetic tunnel junctions (MTJs) provide the framework for current driven and tunable frequency sources with enormous range[1] (from megahertz to gigahertz) that are compatible with existing semiconductor processes. With a direct electrical current applied to the devices, spin-transfer torques transfer angular momentum from a fixed polarizing magnetic layer to a free magnetic layer and induce oscillatory magnetization dynamics.[2,3] The oscillation of the magnetization causes an oscillatory electrical response through the magnetoresistance effect. Due to their small scale, frequency range, and technological compatibility, STNOs may have applications in the telecommunications industry.[4,5] Hurdles to their use come from the large critical current needed to sustain magnetization oscillations with sufficient spectral purity for industrial adoption, as well as low power output. Research has therefore focused on reducing the critical current,[6] decreasing the linewidth,[7] and increasing the power output of STNOs.[6,8] The nonlinearity inherent to STNOs is both the boon and bane of these devices: non-linearities couple the frequency and amplitude of the oscillator, allowing for the large frequency tunability, but also providing the main source of linewidth broadening.[9–11] Experimentally, linewidth reduction has been recently achieved through strategies aimed at controlling the oscillator's phase[12] such as injection locking to an external signal,[13] self-synchronization of several oscillators[14–16] and phase-locked loop techniques.[17]

In this study, we calculate the effect of delayed self-injection on the critical current, frequency response, and linewidth of STNOs. This strategy, where the oscillating output current is re-injected at the

---

input of the oscillator has been shown efficient at improving phase noise in other types of oscillators.[18,19] Using numerical simulations for the dynamics, we find that both the critical current and linewidth of STNOs can be reduced with this technique, while still allowing for frequency tunability. Additionally, we develop simplified analytic expressions that are in good agreement with numerical simulations of the frequency response, critical current, and linewidth – simplifying future experimental and theoretical work. We focus our numerical results on vortex MTJs because they exhibit good output power spectral purity, but emphasize that the analytic results are general to any kind of STNO. The main result is that delayed self-injection technique can be used to decrease critical currents by as much as 25% and linewidths by a factor ≈ 4 for experimentally accessible parameters. We start by describing the model and numerical technique used in this study. We then describe numerical results for the critical current, frequency, and linewidth and compare them with derived analytic expressions.

As illustrated in Fig. 1, the free layer in our system is a magnetic vortex, with a fixed polarizing layer that can have components of its uniform magnetization both in and out-of-plane (in the z-x plane). The resistance of the tunnel junction depends on the core's displacement both radially, $r$, and azimuthally, $\theta$. The overall change in the parallel and anti-parallel components of the vortex magnetic texture relative to the fixed layer contributing to magnetoresistance is zero for displacement of the core along the $x$ direction, and maximal for displacement along the $y$ axis. During oscillation, the junction resistance varies as

$$\Delta R = \lambda \Delta R_0 \rho \sin \theta, \qquad (1)$$

where $\rho = (r/r_0)$, $r_0$ is the disc radius, $\lambda \approx 2/3$ is a geometrical factor[20] describing the amount of vortex magnetic texture parallel/anti-parallel to the polarizer for core displacement to the edge of the nanopillar (along $y$), and $\Delta R_0 = (R_{AP} - R_P)/2$ with $R_P$ and $R_{AP}$ the resistance for parallel and antiparallel alignment of the magnetizations respectively. The delayed self-injected current can be included by adding an oscillating term to the DC current $J_{DC}$, that depends on the history of vortex motion. The effective driving current through the junction when delayed self-injection is included is

$$J = J_{DC}[1 + \chi \, \epsilon \rho_\tau \sin \theta_\tau]. \qquad (2)$$

The subscript $\tau$, represents the time shift by $\tau$ (e.g. $\rho_\tau = \rho(t - \tau)$). $\epsilon = \lambda \frac{\Delta R_0}{R_0}$ is the available microwave current generated by the tunnel junction. $\chi$ is a dimensionless parameter representing the fraction of the microwave current re-injected. If there are losses in the delay circuit $\chi < 1$, but the output may be amplified prior to reinjection. We study the response of the system for $\chi$ up to 10. (Shown schematically in Fig. 1(c)). When $\chi \geq 1$, the circuit will be electrically unstable at frequencies $\frac{n}{\tau}$ for integer $n$. This effect is not present in our simulations. In practice, the working frequency of the system should avoid these frequencies for stability.

Delayed self-injection has some key differences compared to using an alternating current (AC) to drive the magnetization dynamics.[13,21] For an AC drive, when the driving frequency is sufficiently close to the fundamental frequency of the oscillator, the system can lock to the driving frequency. The system also becomes robust to noise near the driving frequency. This behavior is due to the nonlinearity of the system, which allows the oscillator to adjust its frequency to the external driving signal (synchronization).[21] In our case, once transients have resolved, the alternating signal is necessarily at the frequency of the vortex motion but may be in- or out-of-phase when injected depending on $\tau$. The response will depend on $\tau$ in a periodic way. Additionally, the amplitude of an externally driven signal is controlled by the user.[13] Here the self-injected signal depends not only on details of the electronics and MTJ magneto-resistance (through $\chi$ and $\epsilon$), but also the radius of gyration. This may be tuned by the amplification of the delayed signal through $\chi$ and the base DC current driving the dynamics.

To describe the motion of the free magnetic layer, we use the well-established Thiele approach[22] – an effective equation of motion that assumes the coupling to other normal-modes can be neglected or integrated out. For the gyrotropic mode of a vortex,[6] the Thiele equation is

$$G\hat{\mathbf{z}} \times \dot{\mathbf{r}} - D\,\dot{\mathbf{r}} - \frac{\partial W}{\partial \mathbf{r}} + \mathbf{F}_{\mathrm{STT}} = \mathbf{0} \tag{3}$$

where $G$ is the gyrovector magnitude, $D = D_0 + D_1|\mathbf{r}|^2$ the damping, $W$ the confinement potential, and $\mathbf{F}_{\mathrm{STT}}$ the spin-transfer force on the vortex core coordinate, $\mathbf{r}$. Introducing the total current flowing through the junction (Eq. (2)) into the Thiele equation gives (see supplementary material)

$$\begin{aligned}
\dot{\rho} &= a\rho - b\rho^3 + \frac{1}{2}\chi\gamma\rho_\tau \sin(\Delta\theta - \phi_0) - \eta_\theta \cos(\theta - \eta_\theta), \\
\dot{\theta} &= \omega_0 + \omega_1\rho^2 + \frac{1}{2}\chi\gamma\frac{\rho_\tau}{\rho}\cos(\Delta\theta - \phi_0) + \eta_\theta \sin(\theta - \eta_\theta).
\end{aligned} \tag{4}$$

In turn, $a$ and $b$ are the linear and nonlinear effective damping coefficients of the oscillator. The precession rate depends on the linear ($\omega_0$) and nonlinear ($\omega_1$) frequency. $\gamma$ is the effective coupling, $\phi_0$ the associated phase shift, and $\Delta\theta = \theta - \theta_\tau$. $\eta_\rho$ and $\eta_\theta$ are the radial and angular thermal fluctuations which we neglect for now and discuss in detail when evaluating the linewidth. Note that all parameters in Eqn. (4) depend on the DC current density only. In order to produce the form of Eq. (4), an averaging procedure[21] was used to focus on slowly varying quantities. The connection between the parameters of Eqn. (3) and Eqn. (4) is straightforward but cumbersome. (See supplementary information.)

The spin-torque acting on the vortex can be decomposed into three terms, $\mathbf{F}_{\mathrm{STT}} = \mathbf{F}_z + \mathbf{F}_x + \mathbf{F}_{\mathrm{FLT}}$. The first two terms describe the damping-like spin-torque due to the out-of-plane and in-plane components[27] of the fixed polarizer magnet. The third term is the field-like torque (FLT) contribution. The out-of-plane component of the STT effectively opposes the intrinsic damping of the vortex core and can lead to auto-oscillations once the critical current is reached. The terms proportional to $\gamma$ are a direct result of the coupling to the re-injected current. Interestingly, the coupling constant $\gamma$ and associated phase shift $\phi_0$ depend directly on the field-like torque $\mathbf{F}_{\mathrm{FLT}}$ and in-plane component of the damping-like torque $\mathbf{F}_x$. While during pure DC current injection these two forces cannot lead to vortex auto-oscillations they have a huge impact on the dynamics when an alternative current is part of the input.

In the absence of delayed-feedback ($\chi = 0$), Eq. (4) is the generic equation for a non-linear auto-oscillator. It is therefore straightforward to extend these results to any kind of delayed-feedback STNO by considering $\rho$ as the dimensionless amplitude and $\theta$ as the precession angle of a single complex dynamical mode coordinate $c = \rho e^{i\theta}$.[23–26]

To examine vortex dynamics under self-injection, we first solve numerically the delay-differential equations defined by Eqs. (2) and (4). We have integrated these equations using a fourth-order Runge-Kutta scheme for the parameters given in Table I, with time step $\Delta t = 0.5$ ns and simulation time $t_{max} = 10^6$ ns after initial transient. Fig. 2(a) shows the fundamental frequency taken from the Fourier transform of the simulated junction resistance versus delay time and DC current. The phase boundary (white) between damped fluctuations and auto-oscillation is clearly modulated by the delayed signal and periodic in $\tau$. As expected, the phase boundary has periodicity of $2\pi/\omega_0$ (the oscillator period) showing that the phase relation between the re-injected signal and the vortex core position dictates the effective critical current.

In order to gain analytic insight on the critical current reduction, we look for long timescale behavior where, in steady state, the angle grows linearly with time ($\theta = \Omega t$) and the radius becomes fixed ($\dot\rho = 0$ and $(\rho - \rho_\tau) \to 0$). The steady state orbit has a radius that depends on the delay time and periodicity of the oscillator

$$\rho_S = \sqrt{\frac{a}{b} + \frac{\chi\gamma}{2b}\sin(\Omega\tau - \phi_0)}. \tag{5}$$

Including the explicit dependence on the current in Eq. (5) and solving for the critical value ($\rho_S = 0$), we find that the critical current with delayed-feedback, $J^*$, takes the form

$$J^* = \frac{J_0^*}{1 + \chi\zeta \sin(\Omega\tau - \phi_0)} \tag{6}$$

where $J_0^*$ is the critical current in the absence of delayed-feedback and $\zeta$ the scale of the suppression/enhancement of the critical current. This result confirms that the critical current oscillates with the periodicity of the oscillator as $\tau$ is increased and can be reduced by approximately $\chi\zeta$ (for small $\chi\zeta$). In Fig. 2(a), the analytic phase boundary of Eq. (6) is indistinguishable from the numerically determined boundary. Fig. 2(b) shows that Eq. (6) agrees with the simulated critical current for a large range of $\chi$ and suggests critical current suppression by 25 % for large amplification.

Once in steady state, the vortex core radius remains constant and only phase information remains. When $\chi = 0$, Eq. (4) can be solved exactly and previous work has shown a useful definition of the phase is given by $\psi = \theta + \nu \ln \rho$.[21] Here, $\nu = \frac{\omega_1}{b}$ is the nonlinear coupling constant. $\psi$ is chosen to define the dynamics deep in the oscillating regime because it formally grows linearly with time within some neighborhood of the steady state orbit $\rho_S$ even when the radius $\rho$ fluctuates. To make analytic progress, we treat terms proportional to $\gamma$ in Eq. (4) as perturbations. This is justified *a posteriori* by comparing with the simulated response with amplified feedback. We find,

$$\dot\psi = \omega_{NL} + \frac{1}{2}\chi\gamma_\nu \cos(\Delta\psi - \phi_\nu), \tag{7}$$

where $\omega_{NL} = \omega_0 + a\,\nu$ is the unperturbed nonlinear frequency of the system. $\gamma_\nu = \gamma\sqrt{1 + \nu^2}$ is the scaled nonlinear coupling frequency and $\phi_\nu = \phi_0 + \tan^{-1}\nu$ the phase shift with delayed-feedback. In Eq. (7) we have replaced the slowly varying quantity $\Delta\theta$ by $\Delta\psi = \psi - \psi_\tau$ and assumed that $\rho = \rho_\tau$. If we assume the phase grows linearly with time ($\psi \approx \Omega t$) in Eq. (7), we find a transcendental equation for the frequency of the system

$$\Omega = \omega_{\text{NL}} + \frac{1}{2}\chi\gamma_v \cos(\Omega\tau - \phi_v). \tag{8}$$

For small values of the coupling or short delay times, Eq. (8) has one solution bounded between $\omega_{\text{NL}} - \chi\gamma_v/2$ and $\omega_{\text{NL}} + \chi\gamma_v/2$. As the coupling or $\tau$ increase, Eq. (8) has multiple solutions for the frequency whenever $\frac{1}{2}\chi\gamma_v\tau > 1$. Some of these solutions are stable and others are not. To make analytic progress, we define the fluctuation as $\delta\psi = \psi - \Omega t$ and make the assumption that fluctuations are irrelevant after times longer than $\tau$. Then the fluctuations are suppressed provided $\sin(\Omega\tau - \phi_v) > 0$. We find this gives a good estimate of stability. Notice that as $\tau$ increases from a region where a single solution is possible to a region with multiple solutions, the stability analysis suggests there can be a discontinuous jump in the frequency of the system across an unstable region, as seen in the upper-right portion of Fig. 2(a).

With approximate solutions for the critical current, frequency and stability of the oscillator with delayed-feedback we now attempt a description of the system when thermal fluctuations are present. For a rigid vortex magnetic texture the effect of thermal fluctuations can be approximated by a fluctuating magnetic field acting in the Thiele equation. Fig. 3(a) shows the linewidth simulated for $\delta$ – correlated Gaussian white noise with ensemble averages given by

$$\langle \eta \rangle = 0, \qquad \langle \eta_i \eta_j' \rangle = \Gamma\delta_{\text{ij}}\delta(t - t'). \tag{9}$$

$\eta_i(\eta_i')$ is the fluctuating field along Cartesian coordinate $i$ at time $t$ ($t'$) and $\Gamma = 2k_\text{B}T\frac{D_0}{r_0^2 G^2}$ gives the amplitude of fluctuations necessary to maintain thermal equilibrium at temperature $T$ with linear effective damping $D_0$. This approach has been applied to experimental measures of phase and amplitude noise in vortex MTJs with excellent agreement.[28] We have simulated the Thiele equation with this definition of thermal noise in the presence of delayed-feedback.

We evaluate the spectral quality of oscillations by calculating the full-width at half-maximum of the primary spectral peak of the junction resistance at $T = 300$ K (Fig. 3(a)) rather than fitting to a line-shape because we have no *a priori* expectation of the line-shape. Including radial and angular fluctuations in Eqn. (1) gives qualitative features of the resistance variation. We find

$$\langle \delta R^2 \rangle = \frac{(\lambda \Delta R_0)^2}{2}(\langle \delta\rho^2 \rangle + \rho_S^2 \langle \delta\theta^2 \rangle) \tag{10}$$

after averaging over one period. In the absence of the delayed signal, as the trivial state ($\rho_S = 0$) of the vortex is pushed towards steady oscillatory behavior, the linewidth decreases until the critical current is reached. Near the critical current ($\rho_S \neq 0$) both radial and phase fluctuations have a significant effect on the linewidth (full-width at half-maximum). Deep in the oscillatory regime, the linewidth again decreases as radial fluctuations become less relevant and the spectrum is dominated by phase noise.[26] With delayed-feedback (Fig. 3(a)), we see similar trends accompanying the expected oscillatory behavior with delay time. Deep in the oscillatory regime, delayed-feedback can have a dramatic effect on linewidth – either decreasing it (by approximately a factor of 4) or increasing it (by more than a factor of 10).

In the oscillatory regime, it is possible to derive approximate expressions for the linewidth when delayed-feedback is present. Following the derivation of Eq. (7), but including the fluctuating field gives the phase equation

$$\dot{\psi} = \omega_{\text{NL}} + \frac{1}{2}\chi\gamma_v \cos(\Delta\psi - \phi_v) + \frac{\sqrt{1+\nu^2}}{\rho_S}\eta_\rho \sin(\psi - \eta_\theta). \tag{11}$$

The strength of fluctuations in this expression (in polar coordinates) for the phase naturally express the nonlinear broadening of the running frequency in the factor $\sqrt{1+\nu^2}$. The ensemble average of the norm-square of the Fourier transform of Eq. (11) gives the phase noise spectral density in reference to the carrier frequency. This gives

$$S_{\delta\psi}(f) = \langle|\delta\tilde{\psi}|^2\rangle$$
$$= \frac{2\pi\Delta f_0(1+\nu^2)}{\left(2\pi f - \frac{1}{2}\chi\gamma_v\sin(\Omega\tau - \phi_v)\sin 2\pi f\tau\right)^2 + \left(\frac{1}{2}\chi\gamma_v\sin(\Omega\tau - \phi_v)[1 - \cos 2\pi f\tau]\right)^2}, \tag{12}$$

where $\Delta f_0 = \Gamma/2\pi\rho_s^2$ is the linewidth in the absence of nonlinearity and delayed self-injection. Near the carrier frequency ($f \to 0$), the noise spectral density characterizes the linewidth

$$\Delta f = \lim_{\omega \to 0} 2\pi f^2 S_{\delta\psi}(f) = \frac{\Delta f_0(1+\nu^2)}{\left(1 - \frac{\tau}{2}\chi\gamma_v\sin(\Omega\tau - \phi_v)\right)^2}. \tag{13}$$

Eq. (13) agrees with simulated linewidths for large current density (Fig. 3(b)). It suggests that the linewidth can be suppressed by increasing the amplification and delay of the delayed signal. While this is seen clearly in Fig. 3(a), this is the same limit where multiple frequencies may be stabilized. Using the condition of multiple solutions ($\frac{1}{2}\chi\gamma_v\tau \leq 1$) as an upper bound in both the amplification and delay time, Eq. (13) predicts linewidth suppression by a factor of 4 in good agreement with simulation. Complications of working in the regime $\frac{1}{2}\chi\gamma_v\tau > 1$ include the development of sidebands and mode-hopping. (See supplementary information.)

In conclusion, we theoretically investigate the effect of delayed self-injection on critical current, frequency response, and linewidth of STNOs and find that this technique can be used for both critical current and linewidth reduction while maintaining frequency tunability. The dominant coupling derives from the otherwise ineffective field-like and in-plane spin-torques. The importance of this coupling allows for additional design strategies to push STNOs towards commercial constraints. The agreement between our analytic results for critical current, frequency, and linewidth, and the simulated numerical results vortex MTJs highlights the generality of this approach to all STNOs. Additionally, this work expands possibilities of STNOs to high-dimensional dynamics including ultra-efficient synchronization,[29] the possible occurrence and use of chaotic regimes,[30] and brain-inspired reservoir computing.[31]

**Acknowledgements**

JG would like to acknowledge Vincent Cros, Eva Grimaldi, Romain Lebrun and Sumito Tsunegi for fruitful discussions. JG acknowledges funding from the European Research Council, Grant No. 259068.

Table I – Numerical parameters. $\tilde{J} = J_{DC}/(10^8 A/m^2)$.

| Parameter | Simulation Value | |
|---|---|---|
| $R_P$ | 100 | Ω |
| $R_{AP}$ | 200 | Ω |
| G | $1.14 \times 10^{-13}$ | J/(m²rad) |
| $D_0$ | $2.31 \times 10^{-15}$ | J/(m²rad) |
| $D_1$ | $2.31 \times 10^{-15}$ | J/(m²rad) |
| $r_0$ | $2.75 \times 10^{-7}$ | m |
| a | $-9.19 + 1.69\,\tilde{J}$ | MHz |
| b | $11.5 + 8.41 \times 10^{-3}\,\tilde{J}$ | MHz |
| $\omega_0$ | $455 + 0.823\,\tilde{J}$ | MHz |
| $\omega_1$ | $114 - 0.407\,\tilde{J}$ | MHz |
| $\gamma$ | $0.114\,\tilde{J}$ | MHz |
| $\phi_0$ | 0 | rad |
| $J_0^*$ | $5.44 \times 10^8$ | A/m² |
| $\zeta$ | $3.40 \times 10^{-2}$ | |

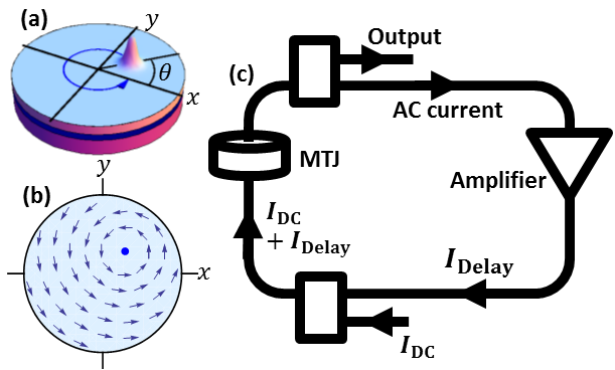

FIG. 1 (Color online) (a) From bottom to top: fixed magnetic layer, insulator, and free vortex magnetic texture. $z$ component of magnetization show schematically near displaced core. (b) In-plane magnetization in the body of the displaced vortex. (c) Schematic circuit diagram of MTJ with delayed-feedback.

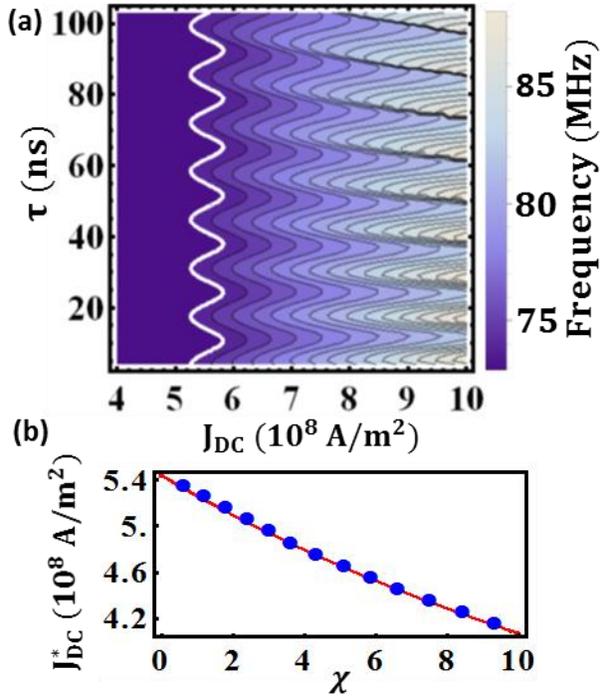

FIG. 2 (Color online) (a) Frequency versus DC current and delay time for $\chi = 1.5$. Critical current for sustained oscillations shown in white. Contours separated by $1.5$ MHz. (b) Critical current versus amplification for $\tau = 18$ ns. Simulated results (blue dots) and the analytic expression (red) shown together.

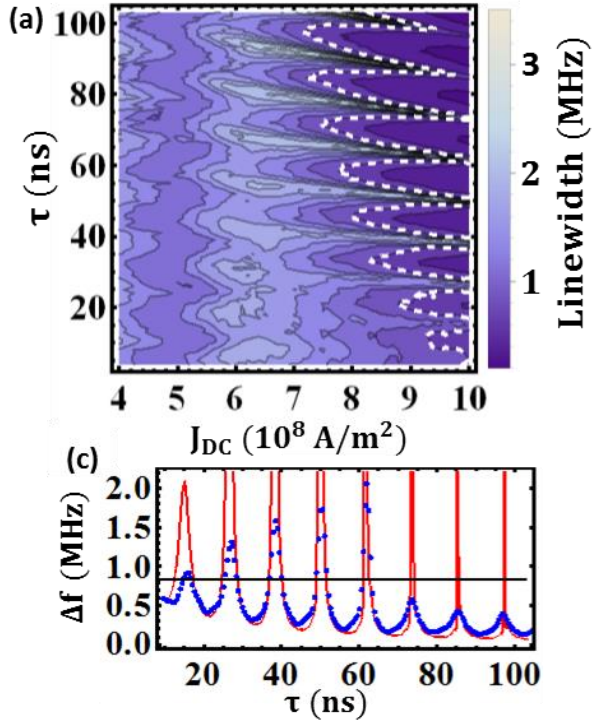

FIG. 3 (Color online) (a) Linewidth variation versus DC current and delay time for $\chi = 1.5$ and $T = 300\ \text{K}$. Contour separation is $1.5\ \text{MHz}$. Lowest linewidth achieved without delayed-feedback shown as white dashed-line. (b) Comparison of simulated linewidth (blue dots) with Eq. (11) - with (red) and without (black) delayed feedback for $J_{DC} = 10^9\ \text{A/m}^2$.

# SUPPLEMENTARY INFORMATION

**Effective Equations with Delayed Self-injection**

Our starting point for describing the dynamics of the free magnetic texture is based on the Thiele approach. In this approach, an effective equation is developed for the response of a magnetic texture to a perturbation. It assumes that the coupling to other spin-modes can be neglected or integrated out so that a simplified description of the spin-wave can be obtained. For the low energy "gyrotropic" mode of vortex free layer, the Thiele equation is given by

$$G\,\hat{\mathbf{z}} \times \dot{\mathbf{r}} - D\dot{\mathbf{r}} - \frac{\partial W}{\partial \mathbf{r}} + \mathbf{F}_{\text{STT}} = 0, \quad (S1)$$

where $\mathbf{r}$ is the coordinate of the vortex core defined in the xy plane. The "gyrovector" points along the z-axis, perpendicular to the nanopillar stack, while the diagonal non-linear damping tensor $D = D_0 + D_1 \frac{|\mathbf{r}|^2}{r_0^2}$ scales the velocity of the vortex core $\dot{\mathbf{r}}$. $r_0$ is the radius of the nanopillar. The third term is the restoring force due to the potential energy gained by shifting the vortex core from equilibrium, and has the general form

$$W(\mathbf{r}) = \frac{1}{2}\kappa(\mathbf{r})|\mathbf{r}|^2 - \mu^*(\hat{\mathbf{z}} \times \mathbf{H}) \cdot \mathbf{r}, \quad (S2)$$

with $\kappa(\mathbf{r}) = \kappa_{\text{MS}}^0 + \frac{1}{2}\kappa_{\text{MS}}^1 \frac{|\mathbf{r}|^2}{r_0^2} + J\kappa_{\text{Oe}}^0 + \frac{1}{2}J\kappa_{\text{Oe}}^1 \frac{|\mathbf{r}|^2}{r_0^2}$. The first two terms in the confinement coupling $\kappa(\mathbf{r})$ describe the linear and non-linear magnetostatic confinement. The third and fourth terms are the Oersted confinement fields when there is a directed current (density) $J$, through the junction. The last term in Eq. (S1) describes the coupling to an external magnetic field through an effective magnetic permeability of the vortex, $\mu^*$. The last term in Eq. (S1) includes all terms related to spin-transfer torque. The fixed magnetic polarizing layer that can have components of its magnetization both in and out-of-plane, taken to lie in the z-x plane. We decompose this nonequilibrium force into three terms, $\mathbf{F}_{\text{STT}} = \mathbf{F}_z + \mathbf{F}_x + \mathbf{F}_{\text{FLT}}$. The first two terms describe the spin-torque due to the out-of-plane and in-plane components of the fixed polarizer magnet. The third term is the field-like torque (FLT) contribution. Together, $\mathbf{F}_x = -a_J^x J\,\hat{\mathbf{x}}$ and $\mathbf{F}_{\text{FLT}} = b_J J\,\hat{\mathbf{y}}$ tend to displace the core away from the center of the pillar. The out-of-plane component of the STT, $\mathbf{F}_z = a_J J\,\hat{\mathbf{z}} \times \mathbf{r}$, effectively opposes the intrinsic damping of the vortex core and can lead to auto-oscillations once the critical current is reached, as described in the main text. The numerical values of the parameters in the Thiele equation have been included in Table S1.

Using the transformations, $\rho = \left|\frac{\mathbf{r}}{r_0}\right|$ and $\theta = \tan^{-1}\left(\frac{\hat{\mathbf{y}}\cdot\mathbf{r}}{\hat{\mathbf{x}}\cdot\mathbf{r}}\right)$ and the definition of the current (Eqn. (1)) leads to the following equations

$$\begin{aligned}\dot{\rho} &= a\rho - b\rho^3 + C_1 + \chi\epsilon\rho_\tau(C_2 + C_3\rho + C_4\rho^3),\\ \dot{\theta} &= \omega_0 + \omega_1\rho^2 + \frac{C_5}{\rho} + \chi\epsilon\frac{\rho_\tau}{\rho}(C_6 + C_7\rho + C_8\rho^2).\end{aligned} \quad (S3)$$

The coefficients of Eqn. (S3) are related to parameters from the Thiele equation (Eqn. (S1)) in Table S2.

We now show that only slowly varying quantities can contribute substantially to the steady state dynamics. Assume that in steady state $\theta = \Omega t$ and $\theta_\tau = \Omega(t - \tau)$. Then terms with the argument $\theta + \theta_\tau = 2\Omega t - \Omega\tau$ vary significantly with time and therefore these terms oscillate and cannot contribute to the long timescale dynamics. Conversely, terms with the argument $\theta - \theta_\tau = \Omega\tau$ are independent of time and can contribute to the long time scale behavior. Therefore, only $C_2$ and $C_6$ can have a sizeable contribution. Using standard trigonometric identities and neglecting terms that vary as $\theta + \theta_\tau$ leads to

$$C_2 \to \frac{1}{2}\frac{b_J}{Gr_0}\sin(\theta - \theta_\tau) - \frac{1}{2}\frac{a_J^x}{Gr_0}\cos(\theta - \theta_\tau) = \frac{1}{2}\frac{\gamma}{\epsilon}\sin(\Delta\theta - \phi_0),$$
$$C_6 \to \frac{1}{2}\frac{a_J^x}{Gr_0}\sin(\theta - \theta_\tau) + \frac{1}{2}\frac{b_J}{Gr_0}\cos(\theta - \theta_\tau) = \frac{1}{2}\frac{\gamma}{\epsilon}\cos(\Delta\theta - \phi_0),$$
(S4)

Where $\gamma = \frac{\epsilon J_{DC}}{Gr_0}\sqrt{b_J^2 + a_J^{x2}}$ and $\phi_0 = \tan^{-1}\left(\frac{a_J^x}{b_J}\right)$. Finally, the averaged equations for the STNO with delayed self-injection are

$$\dot{\rho} = a\rho - b\rho^3 + \frac{1}{2}\chi\gamma\rho_\tau \sin(\Delta\theta - \phi_0),$$
$$\dot{\theta} = \omega_0 + \omega_1\rho^2 + \frac{1}{2}\chi\gamma\frac{\rho_\tau}{\rho}\cos(\Delta\theta - \phi_0).$$
(S5)

Thermal noise can be added straightforwardly by including the fluctuating components of the coefficients $C_1$ and $C_5$ as well. We find,

$$\dot{\rho} = a\rho - b\rho^3 + \frac{1}{2}\chi\gamma\rho_\tau \sin(\Delta\theta - \phi_0) - \eta_\rho \cos(\theta - \eta_\theta),$$
$$\dot{\theta} = \omega_0 + \omega_1\rho^2 + \frac{1}{2}\chi\gamma\frac{\rho_\tau}{\rho}\cos(\Delta\theta - \phi_0) + \frac{\eta_\rho}{\rho}\sin(\theta - \eta_\theta).$$
(S6)

**Derivation of Phase Equation**

Using Eq. (S5) as a starting point and treating all terms proportional to $\gamma$ and the fluctuating magnetic field $\eta$ as perturbations, the phase can be expanded as

$$\dot{\psi} = \partial_\theta\psi\,\dot{\theta} + \partial_\rho\psi\,\dot{\rho}.$$
(S6)

From the phase relation, $\psi = \theta + \nu \ln \rho$, we see that, $\partial_\theta\psi = 1$ and $\partial_\rho\psi = \nu\frac{1}{\rho}$. Therefore,

$$\dot{\psi} = \omega_{NL} + \frac{1}{2}\chi\gamma_\nu \cos(\Delta\psi - \phi_\nu) + \frac{\sqrt{1+\nu^2}}{\rho_0}\eta_\rho \sin(\theta - \eta_\theta - \tan^{-1}\nu)$$
(S7)

using standard trigonometric identities. As in the main text, $\omega_{NL} = \omega_0 + a\,\nu$, $\gamma_\nu = \gamma\sqrt{1+\nu^2}$, and $\phi_\nu = \phi_0 + \tan^{-1}\nu$. The polar thermal fluctuations are related to the Cartesian fluctuations by $\eta_x = \eta_\rho \cos\eta_\theta$ and $\eta_y = \eta_\rho \sin\eta_\theta$.

**Development of sidebands and mode-hopping**

When the delay time and amplification become large enough ($\frac{1}{2}\chi\gamma_\nu\tau > 1$), along with linewidth reduction of the primary peak, the frequency spectrum of the STNO with delayed self-injection develops side-bands. This evolution is shown for a stable regime with large delay time where, as the linewidth decreases, sidebands develop in the frequency spectrum (Fig 3a). In practice these side bands can be removed by further filtering of the output signal.

The largest linewidths found in this simulation occur in regions where the fundamental frequency can be discontinuous (see FIG. 2 of main text). In this region, a large enough perturbation (thermal or otherwise) can induce hopping between neighboring stable frequencies. This provides an additional source of linewidth broadening in FIG. 3a due to the presence of two separated stable frequencies that can be coupled by thermal fluctuations (mode hopping). Figure S2 shows the evolution of such a peak spectrum as temperature is increased. As the temperature is increased a single peak eventually mode-hopping between the two stable frequencies is present.

| Parameter | Simulation Value | |
|---|---|---|
| $G$ | $1.14 \times 10^{-13}$ | J/(m²rad) |
| $D_0$ | $2.31 \times 10^{-15}$ | J/(m²rad) |
| $D_1$ | $2.31 \times 10^{-15}$ | J/(m²rad) |
| $\kappa_{MS}^0$ | $5.20 \times 10^{-5}$ | J/m² |
| $\kappa_{MS}^1$ | $1.30 \times 10^{-6}$ | J/m² |
| $\kappa_{Oe}^0$ | $9.40 \times 10^{-16}$ | J/A |
| $\kappa_{Oe}^1$ | $-4.64 \times 10^{-16}$ | J/A |
| $a_J$ | $1.95 \times 10^{-15}$ | J/A |
| $a_J^x$ | $0$ | J/A |
| $b_J$ | $1.07 \times 10^{-22}$ | J/A |
| $r_0$ | $2.75 \times 10^{-7}$ | m |

Table SI – Numerical parameters for the Thiele equation.

| | | Coefficients of Eq. (S3) |
|---|---|---|
| $\omega_0$ | = | $\frac{1}{G}[\kappa_{MS}^0 + \kappa_{Oe}^0 J_{DC}]$ |
| $\omega_1$ | = | $\frac{1}{G}[\kappa_{MS}^1 + \kappa_{Oe}^1 J_{DC}]$ |
| $a$ | = | $\frac{a_J J_{DC}}{G} - \frac{D_0}{G}\omega_0$ |
| $b$ | = | $\frac{D_1}{G}\omega_0 + \frac{D_0}{G}\omega_1$ |
| $C_1$ | = | $\left(-\eta_x - \frac{b_J J_{DC}}{Gr_0}\right)\cos\theta + \left(-\eta_y - \frac{a_J^x J_{DC}}{Gr_0}\right)\sin\theta$ |
| $C_2$ | = | $-\frac{b_J J_{DC}}{Gr_0}\cos\theta\sin\theta_\tau - \frac{a_J^x J_{DC}}{Gr_0}\sin\theta\sin\theta_\tau$ |
| $C_3$ | = | $\left(\frac{a_J J_{DC}}{Gr_0} - \frac{D_0 \kappa_{Oe}^0 J_{DC}}{G^2}\right)\sin\theta_\tau$ |
| $C_4$ | = | $\left(-\frac{D_0 \kappa_{Oe}^1 J_{DC}}{G^2} - \frac{D_1 \kappa_{Oe}^0 J_{DC}}{G^2}\right)\sin\theta_\tau$ |
| $C_5$ | = | $\left(\eta_x + \frac{b_J J_{DC}}{Gr_0}\right)\sin\theta + \left(-\eta_y - \frac{a_J^x J_{DC}}{Gr_0}\right)\cos\theta$ |
| $C_6$ | = | $-\frac{a_J^x J_{DC}}{Gr_0}\cos\theta\sin\theta_\tau + \frac{b_J J_{DC}}{Gr_0}\sin\theta\sin\theta_\tau$ |
| $C_7$ | = | $\frac{\kappa_{Oe}^0 J_{DC}}{G}\sin\theta_\tau$ |
| $C_8$ | = | $\frac{\kappa_{Oe}^1 J_{DC}}{G}\sin\theta_\tau$ |

Table SII – Relation between the coefficients of Eqn. (S3) and the Thiele equation (Eqn. S1). The thermally fluctuating magnetic fields $\eta_x$ and $\eta_y$ have been included for convenience.

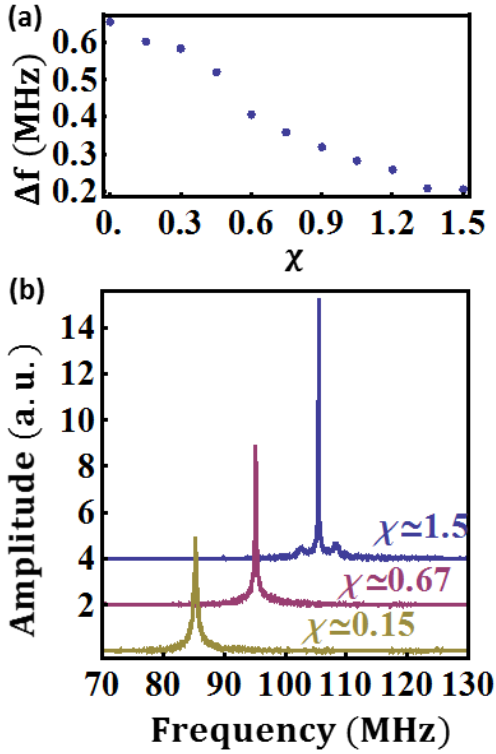

FIG. S1 – Primary peak spectrum variation with amplification. Simulated results shown for $\tau = 90$ ns, $I_{DC} = 10.0 \times 10^8$ A/m$^2$ and $T = 300$ K. (a) Linewidth variation with amplification. (b) As the amplification of the delayed signal is increased, side-bands develop when multiple stable solutions are present. Each spectrum was shifted by 10 MHz laterally from the previous spectrum and uniformly smoothed over 32 kHz for visual clarity.

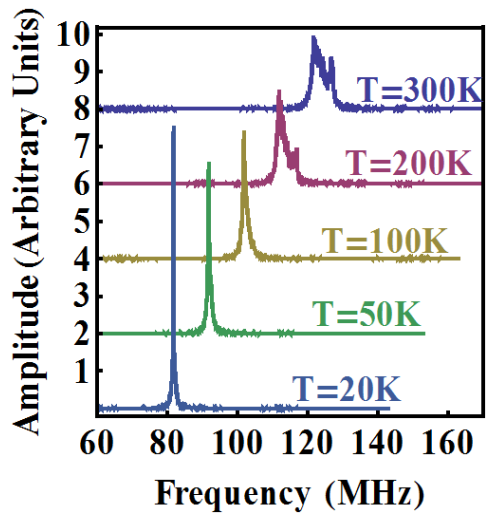

FIG. S2 – Primary peak spectrum near stable discontinuous frequencies. As the temperature is increased, thermal fluctuations induce transitions between stable frequencies near 81 MHz and 87 MHz. Spectrum shown for $\tau = 74$ ns, $I_{DC} = 9.6 \times 10^8$ A/m² and $\chi = 1.5$. Each spectrum was laterally shifted by 10 MHz from the previous spectrum and uniformly smoothed over 32 kHz for visual clarity.